\documentclass{osa-article}

\journal{oe}

\usepackage{graphicx}
\usepackage{amsmath,graphicx,amssymb,braket,xcolor,subfigure,upgreek}
\usepackage{microtype}
\usepackage{bbm}
\usepackage{mathtools}
\usepackage{braket}
\usepackage{placeins}
\usepackage{appendix}
\usepackage{etoolbox}
\apptocmd{\appendices}{\apptocmd{\thesection}{: }{}{}}{}{}

\newcommand{\fref}[1]{Fig.~\ref{#1}}
\newcommand{\ts}[1]{_\text{#1}}
\newcommand{\tr}{\text{tr}}
\newcommand{\blue}[1]{\textcolor{blue}{#1}}
\renewcommand{\Re}[1]{\text{Re}\left\{#1\right\}}
\renewcommand{\Im}[1]{\text{Im}\left\{#1\right\}}
\usepackage{letltxmacro}
\LetLtxMacro{\originaleqref}{\eqref}
\renewcommand{\eqref}{Eq.~\originaleqref}

\begin{document}
\title{Superradiant cooling, trapping, and lasing of dipole-interacting clock atoms}

\author{Christoph Hotter$^*$, David Plankensteiner,\\Laurin Ostermann and Helmut Ritsch}

\address{Institut f\"ur Theoretische Physik, Universit\"at Innsbruck, Technikerstra\ss e 21a\\A-6020 Innsbruck, Austria\\$^*$\blue{\underline{christoph.hotter@uibk.ac.at}}}


\begin{abstract}
A cold atomic gas with an inverted population on a transition coupled to a field mode of an optical resonator constitutes a generic model of a laser. For quasi-continuous operation, external pumping, trapping and cooling of the atoms is required to confine them in order to achieve enough gain inside the resonator. As inverted atoms are high-field seekers in blue detuned light fields, tuning the cavity mode to the blue side of the atomic gain transition allows for combining lasing with stimulated cavity cooling and dipole trapping of the atoms at the antinodes of the laser field. We study such a configuration using a semiclassical description of particle motion along the cavity axis. In extension of earlier work we include free space atomic and cavity decay as well as atomic dipole-dipole interactions and their corresponding forces. We show that for a proper choice of parameters even in the bad cavity limit the atoms can create a sufficiently strong field inside the resonator such that they are trapped and cooled via the superradiant lasing action with less than one photon on average inside the cavity.
\end{abstract}


\section{Introduction}
The idea of building a superradiant laser operating on an ultra-narrow optical clock transition in a cold gas has fostered the vision of implementing the optical analog of microwave clock masers~\cite{vanier1989quantum} with a precision and accuracy improved by many orders of magnitude~\cite{haake1993superradiant,bohnet2012steady,maier2014superradiant,norcia2018cavity}. Today, a central limitation of the best optical clock implementations~\cite{campbell2017fermi,leopardi2018absolute} is noise within the mirrors of the reference oscillators~\cite{ludlow2007compact} that act as the flywheels locked to the atomic transition frequency. When operated on the clock transition in the bad cavity regime and at low photon numbers, superradiant lasers have been predicted to be very insensitive to these fluctuations and create an accurate and precise frequency reference~\cite{meiser2010steady,bohnet2012steady, maier2014superradiant,norcia2016superradiance}.

In principle, operated at high photon numbers sufficiently above the lasing threshold, lasers do not exhibit a fundamental limit of their linewidth~\cite{schawlow1958infrared,kuppens1994quantum}. In practice, however, the operational laser linewidth is determined by technical noise in the resonator and in the active medium. Technological advances have reduced this limit down to the order of Hz~\cite{kessler2012sub}, which has lead to a growing interest in using long-lived clock states as the gain medium in a new generation of so-called superradiant lasers~\cite{meiser2009prospects,bohnet2012steady,bohnet2013active,vuletic2012an_almost}. However, the long lifetimes, i.e.\ the small linewidths, of those states entail a minute dipole moment of the involved transitions, thus making it necessary to work in the strong collective coupling regime. In this domain, by means of synchronization through the cavity field~\cite{weiner2017phase,xu2013simulating,henschel2010cavity}, a large collective dipole will build up, which can provide the necessary gain. Here, the atoms do not need to be confined in a small volume as is the case with Dicke superradiance~\cite{dicke1954coherence}, but they have to couple almost equally to the cavity.

In the present manuscript we investigate a model of a superradiant laser where the gain medium is self-trapped by the cavity field it creates via stimulated emission into the resonator. At a suitably chosen detuning of the cavity above the atomic transition frequency, the atoms will also slow down and experience a cooling within their prescribed trap positions while simultaneously acting as the gain medium for the laser~\cite{salzburger2004atomic,salzburger2006lasing}. Recently, very efficient cooling has been predicted involving cavity mediated collective superradiant decay and atomic dipole-dipole interactions~\cite{xu2016supercooling}. Since inverted atoms in a blue detuned cavity are high-field seekers they are drawn to mode antinodes and almost equal coupling can be achieved.

\section{Model}
\begin{figure}[ht]
\centering
\includegraphics[width=0.8\columnwidth]{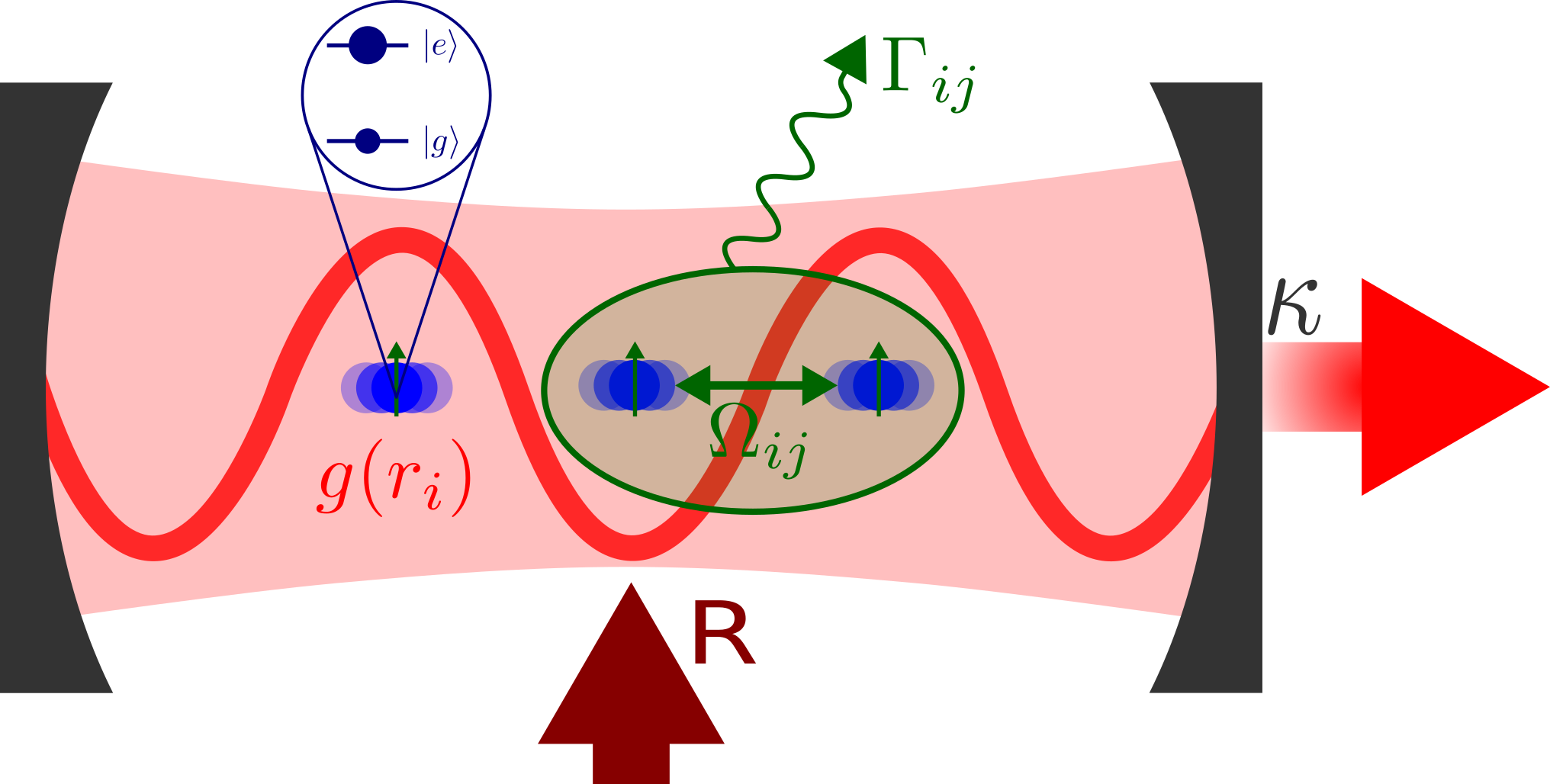}
\caption{\emph{Schematic of our model.} We consider a confined ensemble of two-level atoms moving along the axis of a cavity resonantly coupled to a single mode with amplitude $g(r_i)$. The atoms directly interact via resonant dipole-dipole coupling inducing pairwise energy exchange $\Omega_{ij}$ and collective decay with decay rates $\Gamma_{ij}$. A uniform transverse pump mechanism individually excites atoms at rate $R$, while the cavity loses photons at rate $\kappa$.}
\label{fig:model}
\end{figure}

Let us consider $N$ identical two-level atoms confined to one-dimensional motion along the axis of a Fabry Perot cavity. At finite temperature we can assume a classical description of atomic motion along the cavity axis. All transition dipoles are assumed parallel and perpendicular to the cavity axis as in a $J=0 \to J=1$ transition. The atoms couple to the cavity mode via the well known Tavis-Cummings interaction with a strength given by the cavity mode function at the atomic position, $g(r_i)$. Given that the atomic ensemble is closely spaced, we need to take coherent dipole-dipole energy exchange ($\Omega_{ij}$) and collective spontaneous emission ($\Gamma_{ij}$), which are both mediated by the surrounding vacuum, into account. Furthermore, we assume to create population inversion of the relevant two atomic levels via an individual transverse incoherent pump with the rate $R$. In practise this has to be implemented via a multistep process involving intermediate levels. Photons can leak through the cavity mirrors at a cavity loss rate $\kappa$ (see \fref{fig:model}).

The Hamiltonian of this system in the rotating wave approximation and in the reference frame of the atoms is
\begin{equation}
H = \hbar\Delta a^{\dagger} a + \sum_{i = 1}^{N} \hbar g(r_i) [ a \sigma_{i}^{+} + a^{\dagger} \sigma_{i}^{-}] + \sum_{i,j : i \neq j} \hbar\Omega_{ij}\sigma_{i}^{+}\sigma_{j}^{-},
\label{Hamiltonian}
\end{equation}
where $a^\dagger$ ($a$) is the bosonic creation (annihilation) operator which creates (annihilates) a photon with frequency $\omega\ts{c}$ in the cavity. The operators $\sigma_i^+$ and $\sigma_i^-$ are the atomic raising and lowering operators of the $i$th two-level atom with transition frequency $\omega\ts{a}$. The $i$th dipole couples to the cavity mode with the position-dependent coupling strength $g(r_i) = g \cos(k\ts{c} r_i)$. The coupling constant is denoted by $g$ and $k\ts{c} = 2 \pi / \lambda\ts{c}$ is the wave number of the cavity mode. The frequency $\Omega_{ij}$ quantifies the resonant dipole-dipole energy transfer between atoms $i$ and $j$. The detuning between the cavity resonance frequency and the atomic transition frequency is given by $\Delta = \omega\ts{c} - \omega\ts{a}$.

Dissipative processes are accounted for by the Liouvillian $\mathcal{L}$ in the master equation
\begin{equation}
\dot{\rho} = - \frac{i}{\hbar} \left[ H,\rho \right] + \mathcal{L}\left[\rho\right].
\label{eq:master_eq}
\end{equation}
Within the Markov approximation our Liouvillian consists of three parts, namely
\begin{equation}
\mathcal{L}[\rho] = \mathcal{L}_{\mathrm{pump}}[\rho] + \mathcal{L}_{\mathrm{cav}}[\rho] + \mathcal{L}_{\mathrm{cd}}[\rho],
\label{eq:Liouvillian}
\end{equation}
where the individual incoherent transversal pump is characterized by the pump rate $R$,
\begin{equation}
\mathcal{L}_{\mathrm{pump}}[\rho] = \frac{R}{2} \sum_i (2 \sigma_i^+ \rho \sigma_i^- -  \sigma_i^- \sigma_i^+ \rho - \rho \sigma_i^-\sigma_i^+),
\label{eq:Liouvillian_pump}
\end{equation}
the cavity losses occur at the cavity decay rate $\kappa$,
\begin{equation}
\mathcal{L}_{\mathrm{cav}}[\rho] = \kappa (2 a \rho a^{\dagger} - a^{\dagger} a \rho - \rho a^{\dagger} a),
\label{eq:Liouvillian_cavity}
\end{equation}
and the collective atomic decay is determined by the generalized spontaneous emission rates $\Gamma_{ij}$,
\begin{equation}
\mathcal{L}_{\mathrm{cd}}[\rho] = \frac{1}{2} \sum_{ij} \Gamma_{ij} (2 \sigma_i^- \rho \sigma_j^+ -  \sigma_i^+ \sigma_j^- \rho - \rho \sigma_i^+ \sigma_j^-).
\label{eq:Liouvillian_cd}
\end{equation}

The resonant dipole-dipole couplings $\Omega_{ij}$ and the collective decay rates $\Gamma_{ij}$ depend on the interatomic distances~\cite{lehmberg1970radiation,ficek2002entangled} and are given by
\begin{equation}\label{eq:Omega_ij}
\Omega_{ij} = - \frac{3 \Gamma}{4} \Big[(1-\cos^2 \Theta) \frac{\cos(k_a r_{ij})}{k_a r_{ij}} - (1-3 \cos^2\Theta) \left(\frac{\sin(k_a r_{ij})}{(k_a r_{ij})^2} + \frac{\cos(k_a r_{ij})}{(k_a r_{ij})^3}\right)\Big] ,
\end{equation}
and
\begin{equation}\label{eq:Gamma_ij}
\Gamma_{ij} = \frac{3 \Gamma}{2} \Big[(1-\cos^2\Theta) \frac{\sin(k_a r_{ij})}{k_a r_{ij}} + (1-3 \cos^2\Theta) \left(\frac{\cos(k_a r_{ij})}{(k_a r_{ij})^2} - \frac{\sin(k_a r_{ij})}{(k_a r_{ij})^3}\right)\Big].
\end{equation}
Here, $k\ts{a} = \omega\ts{a}/c$ is the wavenumber corresponding to the atomic transition frequency and $\Theta$ denotes the angle between the atomic dipoles and the distance vector between atom $i$ and atom $j$.

For the time evolution of the classical variables we have for the velocity of the $i$th particle
\begin{equation}
\dot{r}_i = \frac{p_i}{m} = 2 \omega\ts{r} \frac{p_i}{\hbar k\ts{a}^2},
\label{ehrenfest_dr}
\end{equation}
and the force acting on a particle is (see Appendix~\ref{app:semi_class_approx} for details)
\begin{equation}
\dot{p}_i = - \hbar\partial_{r_i} \bigg[ g(r_i) \braket{a \sigma_{i}^{+} + a^{\dagger} \sigma_{i}^{-}} + \sum_{j:j \neq i} 2 \Omega_{ij}\Re{\braket{\sigma_{i}^{+}\sigma_{j}^{-}}} \bigg].
\label{ehrenfest_dp}
\end{equation}
Here, we defined $\omega\ts{r} := \hbar k\ts{a}^2 / (2m)$ as the recoil frequency, with $m$ the mass of an atom.

Note that the above equations are only valid for sufficiently slow particles. This is because we include the time dependence of the collective dipole-dipole interactions via the time-dependent atomic positions only. Thus, Doppler shifts and other effects depending on the velocity, or higher-order derivatives of the position, are neglected in the dipole-dipole coupling. Furthermore, we note that forces stemming from the collective decay are neglected here (see Appendix~\ref{app:semi_class_approx}). Additionally, since we assume classical motion, the recoil from spontaneous emission is neglected in the kinetic energy. This is probably the most drastic approximation made here.

\section{Cooling and trapping properties} \label{sec:cooling_and_trapping_properties}
We investigate the stability of the system described above in the lasing regime by showing that the atoms are cooled and trapped within the cavity field potential created by the photons scattered from the inverted atoms. Due to the exponential scaling of the Hilbert space dimension with the number of atoms numerical methods are limited. Thus, we restrict ourselves to treating a sufficiently small system that still exhibits collective effects.

We study a system with three atoms inside the cavity. The initial state is as follows. The atoms are placed $\lambda\ts{c}/2$ apart at the cavity field antinodes. Furthermore, the atoms are in the ground state and there are no photons inside the cavity. The set of initial momenta is picked from a normal distribution that depends on the recoil frequency. Namely, we always choose a normal distribution for the atomic momentum such that the average kinetic energy is constant in respect to $\omega\ts{r}$. In general, the kinetic energy of the $i$th particle is $p_i^2/(2m)$. Thus, if the recoil frequency is multiplied by an arbitrary constant $c$ (i.e. the mass is divided by $c$), we need to scale the momentum with $1/\sqrt{c}$ in order to keep the kinetic energy constant. The standard deviation $\bar{p}_0$ of the initial momentum distribution of the atoms is chosen depending on the choice of $\omega\ts{r}$ according to this relation.
\begin{figure}[t]
\centering
\centering\includegraphics[width=\columnwidth]{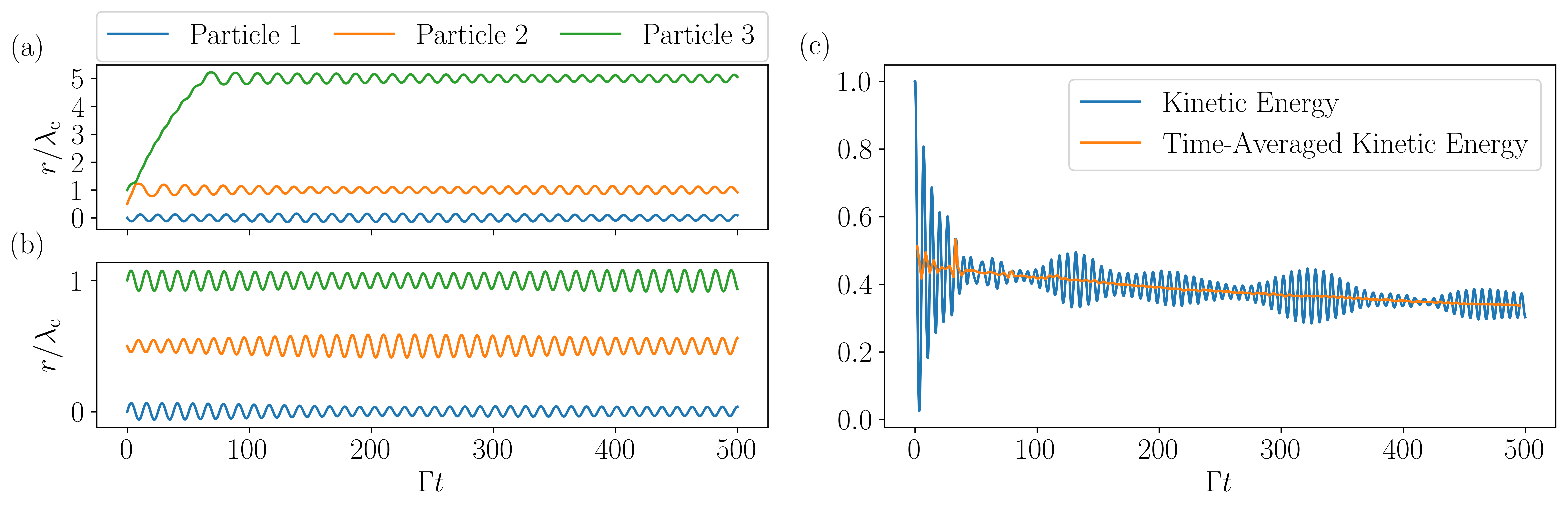}
\caption{\emph{Exemplary trajectories of three particles and their time-averaged kinetic energy loss}. In figure (a) an initially untrapped particle (green and orange line) is slowed down by cavity cooling until it is trapped, whereas all other particles in the examples remain close to their initial trapping position near a field antinode. The set of initial momenta is $p_0 \simeq [-1.78, 3.92, 2.83]\hbar k\ts{a}$ in (a) and $p_0 \simeq [1.00, -0.73, 1.18]\hbar k\ts{a}$ in (b). In (c), we see that the time-averaged relative kinetic energy does not vary that much in time as opposed to the momentary kinetic energy exhibiting trapped oscillatory motion. For better visibility we normalize the kinetic energy to its initial value. The parameters are $N=3$, $\omega\ts{r} = 0.1 \Gamma$, $\Delta = 10\Gamma$, $g = 5\Gamma$, $\kappa = 10\Gamma$ and $R = 8\Gamma$ for all three figures.}
\label{fig:position_ab1_ab22_ekin_average}
\end{figure}

To analyze the trapping and cooling properties of the system we study the time evolution of the particle positions. In \fref{fig:position_ab1_ab22_ekin_average}(a) we show a case where the particles are cooled until they are cold enough to get trapped in the potential created by the cavity field. The particles distribute themselves relatively far from each other, which means that collective effects hardly play a role. As we aim to investigate collective effects as well, we restrict our calculations to particle trajectories that remain in their initially prescribed trap for the entire time evolution. We call them completely stable trajectories, see for example \fref{fig:position_ab1_ab22_ekin_average}(b). We refer to Appendix~\ref{app:stability} for more details on the stability. The momentum transfer from particle 2 to particle 3, depicted in \fref{fig:position_ab1_ab22_ekin_average}(a), stems from the collective dipole-dipole effects.

In order to quantitatively capture the cooling process, we study the time evolution of the particles' kinetic energy
\begin{align}
E\ts{kin}(t) = \sum_i \frac{p_i(t)^2}{2m},
\end{align}
and average over $100$ thermally distributed initial momenta. However, we consider the completely stable trajectories only. As evident from \fref{fig:position_ab1_ab22_ekin_average}(c) (blue line) this kinetic energy of the stable trajectories oscillates very rapidly on the time scale of the cooling process and thus does not yield comparable results. Therefore, we introduce the time-averaged kinetic energy $\bar{E}\ts{kin}(t)$, which is obtained by taking the midpoints between two adjacent extrema of the kinetic energy as seen in \fref{fig:position_ab1_ab22_ekin_average}(c) (orange line). The parameter we use in order to characterize cooling or heating is
\begin{equation}
\bar{E}\ts{kin}^\mathrm{rel}(t) = \frac{\bar{E}\ts{kin}(t)}{\bar{E}\ts{kin}(0)},
\label{eq:e_kin_time-average}
\end{equation}
which we call time-averaged relative kinetic energy.

\begin{figure}[t]
\centering
\includegraphics[width=0.8\columnwidth]{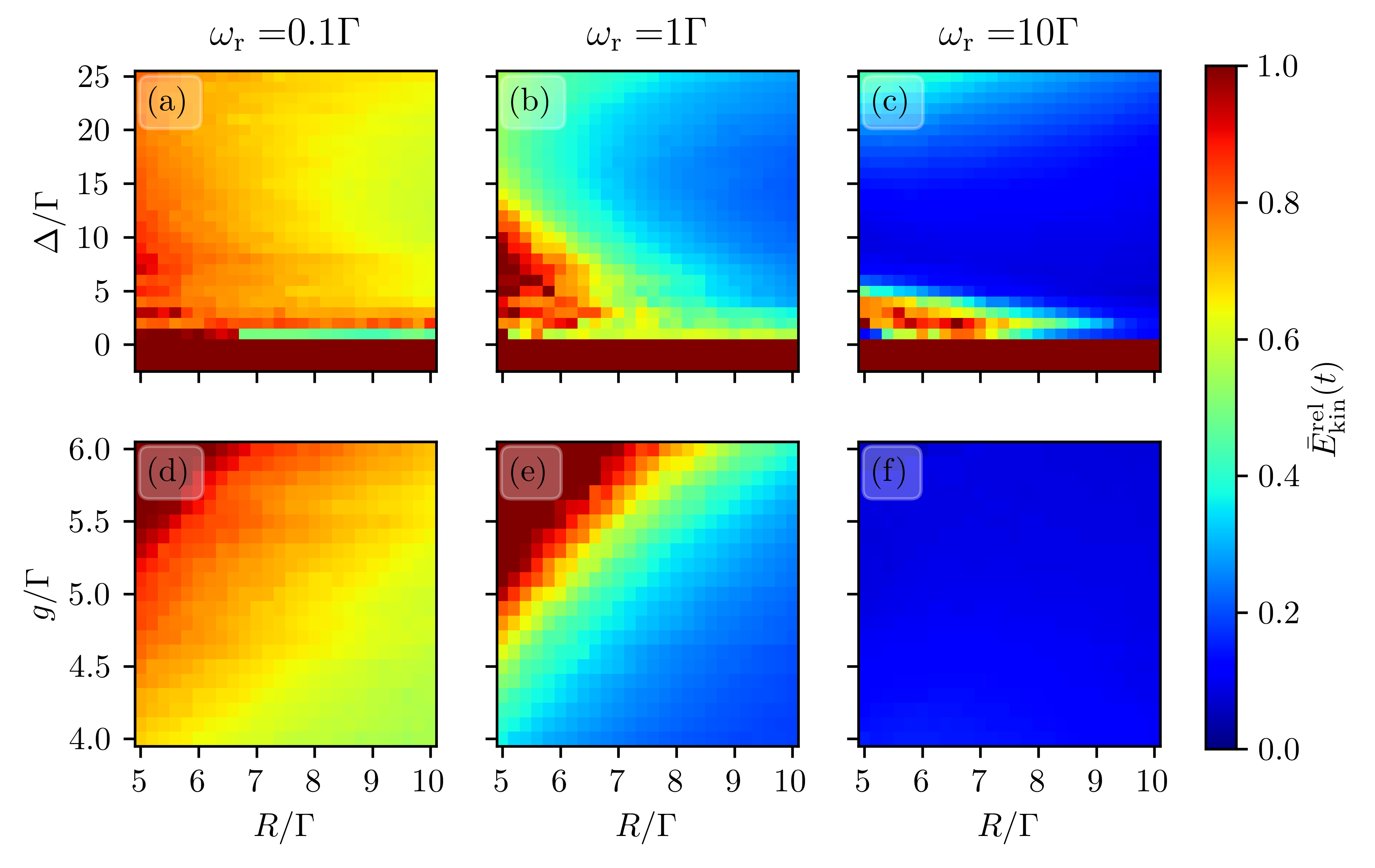}
\caption{\emph{Cycle-averaged relative kinetic energy change after an evolution time of 500 atomic lifetimes as function of various operating parameters.} We show scans over $\Delta$ and $R$ in (a),(b) and (c) and vary $g$ and $R$, respectively, in (d),(e) and (f) for different $\omega\ts{r}$. The parameters when kept constant are $N=3$, $\Delta = 5\Gamma$, $g = 5\Gamma$, and $\kappa = 10\Gamma$ and $t=500/\Gamma$.}
\label{fig:e_kin_rel_scan}
\end{figure}

We scan over the experimentally most accessible parameters using the procedure described above for three different $\omega\ts{r}$. The thermally distributed initial kinetic energy corresponds to a normal distribution of the initial momenta. As discussed above, in order to ensure that the particles start with the same average kinetic energy for all values of $\omega\ts{r}$ we scale the standard deviation $\bar{p}_0$ of the momentum distribution. We choose $\bar{p}_0 = 2\hbar k\ts{a}$ for $\omega\ts{r} = 0.1\Gamma$, $\bar{p}_0 = 2/\sqrt{10} \hbar k\ts{a}$ for $\omega\ts{r} = 1\Gamma$ and $\bar{p}_0 = 2/10 \hbar k\ts{a}$ for $\omega\ts{r} = 10\Gamma$. \fref{fig:e_kin_rel_scan} shows the scan over $\Delta$ and $R$ as well as over $g$ and $R$ for all three values of $\omega\ts{r}$. Every value of $\bar{E}\ts{kin}^\mathrm{rel}(t)$ above $1.0$ corresponds to heating and is artificially fixed to $1.0$, as these are points of little interest.

We observe that only trajectories with $\Delta > 0$ realize cooling, which corresponds to the expected blue detuning of the cavity mode with respect to the atoms. This is due to the fact that atoms inside a cavity favour the emission of photons near the cavity resonance frequency~\cite{ritsch2012cavitypotentials}. An atom in a blue detuned cavity emits photons at a frequency higher than its transition frequency. Therefore, the atom has to exert energy in order to lose a photon into the cavity, which it does by losing kinetic energy. The atoms feel an effective friction force that is largest at the points where the cavity field is maximal (high-field seeking). As can be seen from the scans, there is an optimum for the detuning where the cooling is maximal. This is similar to the maximal force in the process of Doppler cooling. The force is also proportional to the excited state population of the atoms. The cooling is thus best when the pump is sufficiently strong to keep the atoms inverted at almost all times.

Furthermore, we can see that atoms with a larger $\omega\ts{r}$ reach a lower relative kinetic energy during a fixed cooling time. On the one hand, this means that lighter particles are cooled down faster. On the other hand, their initially larger velocities make them more difficult to trap, i.e. more trajectories are unstable (see Appendix~\ref{app:stability}). Heavier atoms (smaller $\omega\ts{r}$) are easier to trap for the same initial kinetic energy (see \fref{fig:stability}), even though they do not cool as much during the observed time interval. Note the difference between cooling and trapping here: heavier atoms can still be trapped if they cool poorly, since even a larger kinetic energy in this case oftentimes corresponds to a relatively small velocity insufficient for the particles to climb the potential walls of the trapping potential created by the cavity field. Since they start with a lower initial velocity, however, the cooling is much slower. The inverse line of argument holds for lighter atoms: they are more difficult to trap, but if they are trapped the cooling is more efficient.

Finally, as can be seen in Figs.~\ref{fig:e_kin_rel_scan}(d)-\ref{fig:e_kin_rel_scan}(f), the coupling to the cavity mode should not be too large in order for the system to cool the atoms. This can be explained by the growing probability of the atoms absorbing photons from the cavity, which causes heating. Hence, the coupling strength should always be well below the cavity loss rate, such that it is much more probable for a photon to leave the cavity than to be reabsorbed. Higher pump strengths can also counteract the heating. If the atoms are pumped strongly they are inverted at almost all times and thus cannot absorb an incident cavity photon. Note that the red areas in \fref{fig:e_kin_rel_scan} with $\Delta > 0$ are mainly caused by extremely slow initial atoms. In these cases the atoms do remain trapped, even though they are heated (note again the difference between cooling and trapping). They are initially so slow that the noise stemming from the cavity field causes heating inside their trap. If the atoms here started with a larger kinetic energy (temperature), they would indeed be cooled. However, they would then also be fast enough to leave their initial traps.

At this point we would like to emphasize again that we describe the system in a semiclassical treatment and we neglect the recoil arising from spontaneous emission. Since the absolute values of the particles' momenta are around $\hbar k\ts{a}$ we need to view the cooling and trapping results critically, especially for $\omega\ts{r} = 1\Gamma$ and $\omega\ts{r} = 10\Gamma$. Note, though, that the recoil heating would only have an effect here since we formulate the superradiant laser regime in terms of a toy model. Specifically, the spontaneous emission rate is taken to be in the limit where $\Gamma\ll\kappa$. However, it is still chosen much larger than it would be for realistic clock atoms in order to avoid numerical difficulties due to different timescales. While we choose $\Gamma \sim 10^{-1}\kappa$, a more realistic choice would be $\Gamma\sim 10^{-6}\kappa$. In that case, a spontaneous emission event is so rare that the recoil can be safely neglected. Still, the fact that we do not include recoil effects for our choice of parameters may be viewed as a rather drastic simplification in our model.

\subsection{Collective cooling effects}
\begin{figure}[ht]
\centering
\centering\includegraphics[width=0.8\columnwidth]{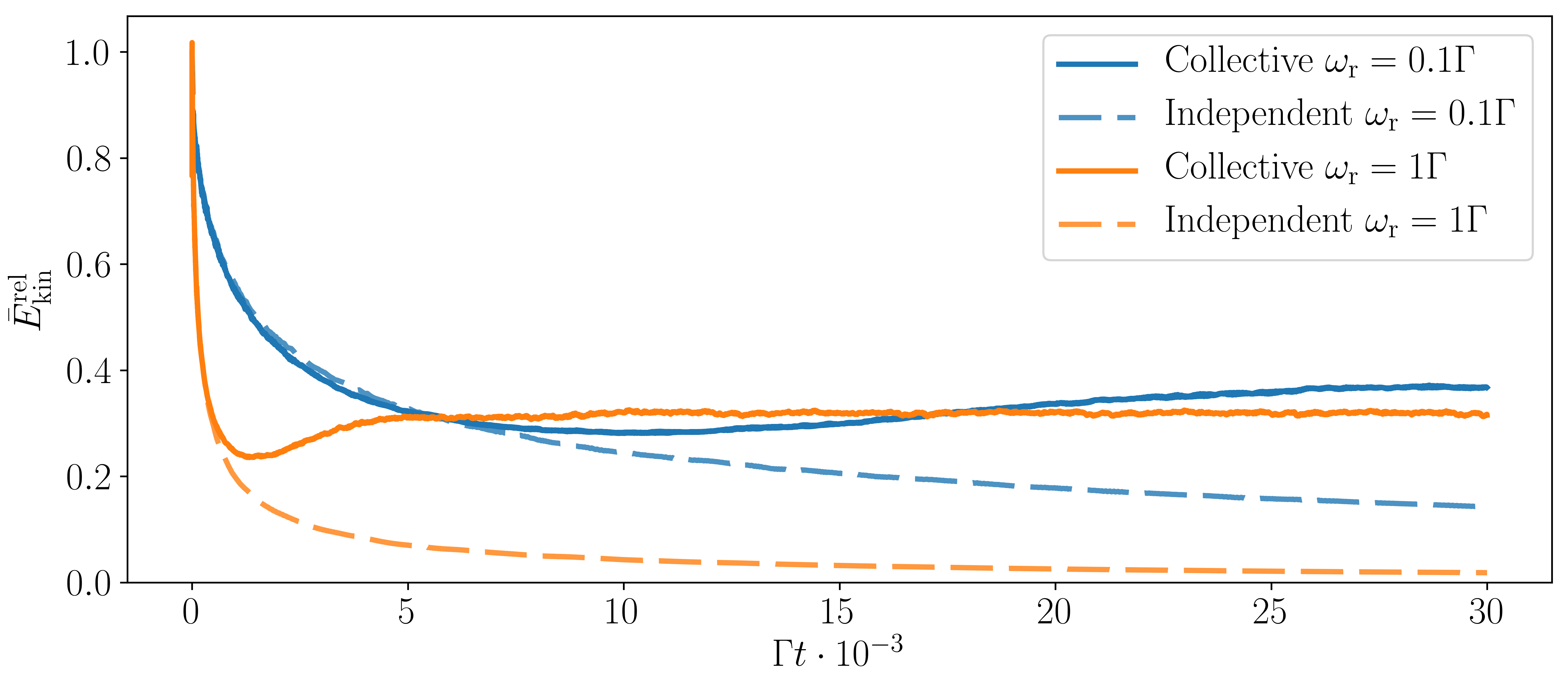}
\caption{\emph{Comparison of motional cooling in time with and without direct dipole interaction}. We compare the motional energy loss for collectively interacting (solid lines) and independent atoms (dashed lines)  showing $\bar{E}\ts{kin}^{\mathrm{rel}}$ for both cases. The independent case describes atoms far apart from each other. The parameters are $N = 3$, $\Delta = 5\Gamma$, $g = 5\Gamma$, $\kappa = 10\Gamma$ and $R = 8\Gamma$ for both, $\omega\ts{r} = 0.1 \Gamma$ and $\omega\ts{r} = 1 \Gamma$.}
\label{fig:coll_vs_indep_cooling}
\end{figure}

Let us now investigate the relevance of the collective effects for the cooling process. Therefore, we set $\Omega_{ij} = 0$ and $\Gamma_{ij} = \delta_{ij} \Gamma$, and compare this independent cooling to the collective cooling from above. In order to acquire collective effects we find that we need to extend the cooling time by orders of magnitude. In \fref{fig:coll_vs_indep_cooling} we see the time evolution of $\bar{E}\ts{kin}^{\mathrm{rel}}$ for collectively interacting atoms in comparison to independent ones.

The main result from \fref{fig:coll_vs_indep_cooling} is that independent atoms will always reach a lower final kinetic energy for long cooling times. In the collective case the atoms push or pull each other away from the cavity field antinodes and thus their displacement amplitude is larger. Therefore, their kinetic energy is bigger on average. Until approximately $\Gamma t = 5000$ the collective line is slightly below the independent line, for $\omega\ts{r} = 0.1 \Gamma$. The reason for this is that parts of the kinetic energy are absorbed into the dipole-dipole interaction potential. The fact that there is a minimum below the final value in the collective case stems also from the dipole-dipole interaction. The minimal temperatures in the two collective cases shown in \fref{fig:coll_vs_indep_cooling} are reached at approximately the same time in units of $\omega\ts{r}$.

As we mentioned in the beginning, we restrict our considerations to $N=3$ due to the exponential growth of the Hilbert space. Let us still comment on what one might expect for a larger number of atoms in terms of cooling. In~\cite{xu2016supercooling}, it has been shown that efficient cavity cooling can be achieved without direct dipole-dipole interactions between the atoms. Rather, the interactions there stem from the cavity mediated dipole coupling. These findings in combination with the result shown in \fref{fig:coll_vs_indep_cooling} suggest that the limit imposed on the final kinetic energy by direct dipole-dipole coupling will be more pronounced for larger atom numbers. More precisely, the cooling without direct dipole-dipole interactions yields lower final kinetic energies with growing atom number~\cite{xu2016supercooling}. The larger energy due to the displacement caused by direct dipole-dipole interactions should thus cause an increasing difference to non-interacting atoms.

\section{Laser properties}
After having established that the system is stable for a given set of parameters, we proceed by analyzing its lasing properties. To this end we study the cavity spectrum, the average photon number as well as the second-order correlation function. Furthermore, we look at the atomic inversion. We use the density matrices at $\Gamma t = 500 $, which describe a quasi-stationary final state, in order to calculate the properties mentioned above.

\subsection{Laser spectrum}  
\begin{figure}[t]
\centering
\centering\includegraphics[width=\columnwidth]{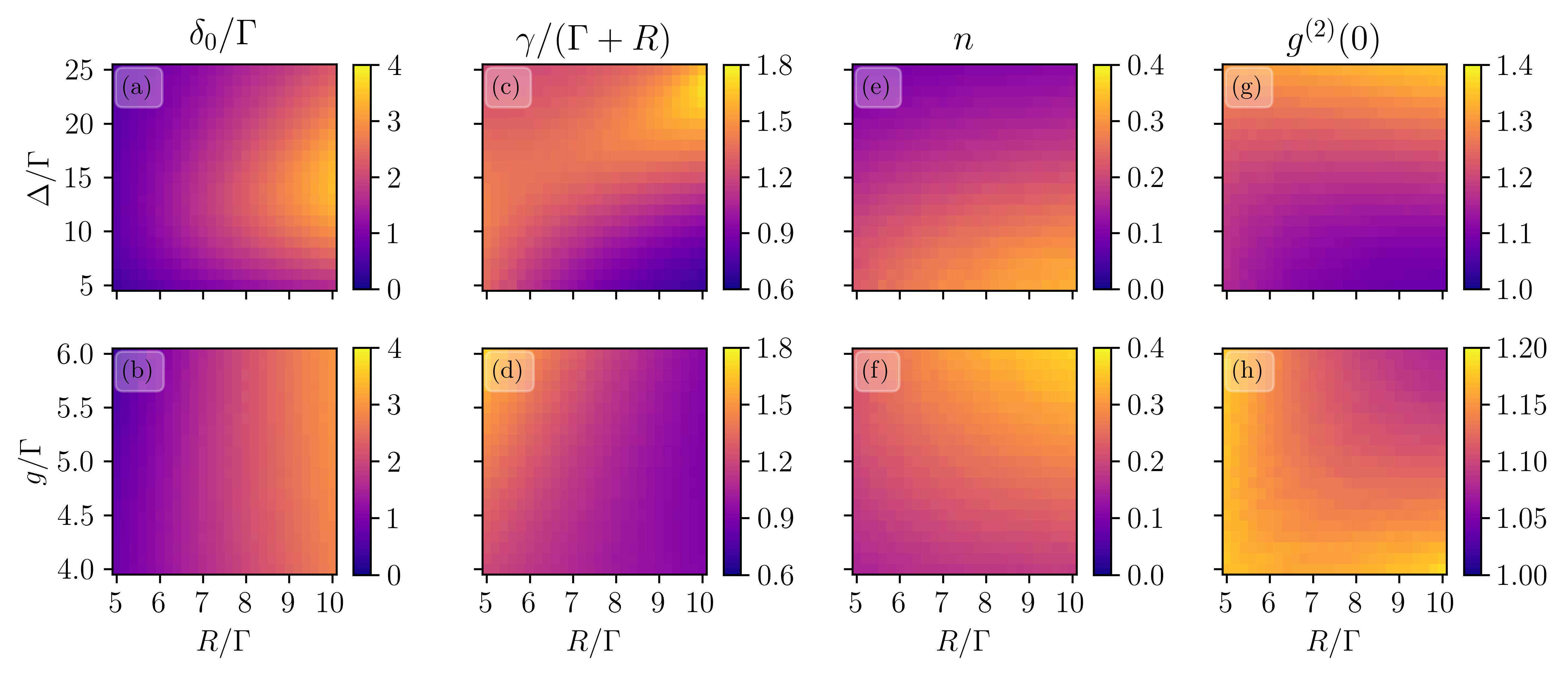}
\caption{\emph{Properties of the emitted laser light as a function of different system parameters}. We depict scans for the peak frequency shift $\delta_0$ from the atomic resonance in (a) and (b), the laser linewidth $\gamma$ in (c) and (d), the average photon number $n$ in (e) and (f), and the bunching parameter $g^{(2)}(0)$ in (g) and (h). We focus on values of $\Delta \geq 5\Gamma$, since below this threshold the particle motion shows few stable trajectories only (see \fref{fig:stability}). The remaining parameters are the same as in \fref{fig:e_kin_rel_scan}.}
\label{fig:lasing_scan}
\end{figure}
The laser spectrum can be calculated as the Fourier transform of the first order correlation function $g^{(1)}(\tau) = \braket{a^{\dagger}(t+\tau)a(t)}$. According to the Wiener-Khinchin theorem~\cite{puri_mathemmathicalmethods}, we have
\begin{equation}
S(\omega) = 2\Re { \int_{0} ^{\infty} \mathrm{d} \tau e^{-i \omega \tau} g^{(1)}(\tau)}.
\label{eq:Wiener-Khinchin}
\end{equation}

\begin{figure}[t]
\centering
\centering\includegraphics[width=0.8\columnwidth]{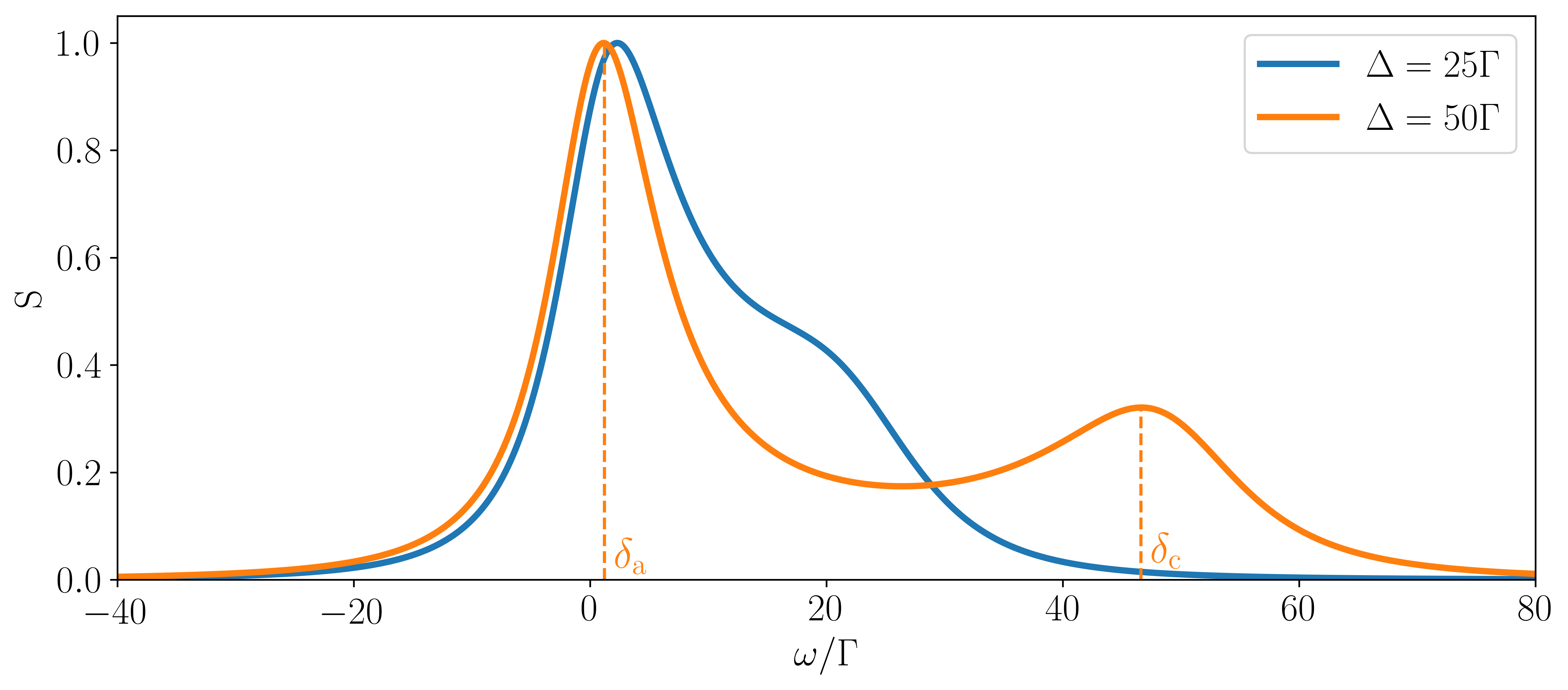}
\caption{\emph{Appearance of a second maximum in the spectrum for large atom cavity detuning}. For detunings larger than $\Delta = 25 \Gamma$ a second peak emerges at the cavity resonance to the right-hand side of the atomic peak. For $\Delta = 50 \Gamma$ this peak is almost completely separated. This shifts the average of the emitted intensity towards the cavity resonance. We call the offset from the atomic transition frequency $\delta\ts{a}$ and the one from the cavity resonance $\delta\ts{c}$. The parameters are $N=3$, $\omega\ts{r} = 0.1 \Gamma$, $g = 5\Gamma$, $\kappa = 10\Gamma$ and $R = 10\Gamma$ for both, $\Delta = 25\Gamma$ and $\Delta = 50\Gamma$.}
\label{fig:spectrum_2peak}
\end{figure}

Appendix~\ref{app:calc_spec} provides details on how the spectrum is calculated in our semiclassical approximation. Most of the spectra are well described by a Lorentzian distribution  (see \fref{fig:spectrum_fit}). Thus, we determine the full width at half maximum (FWHM) $\gamma$ and the offset to the atomic resonance frequency $\delta_0$. The dependency of the linewidth $\gamma$ and the offset $\delta_0$ on our scan parameters is depicted in Figs.~\ref{fig:lasing_scan}(a)-\ref{fig:lasing_scan}(d). We show these plots for one choice of the recoil frequency only, namely $\omega\ts{r} = 1 \Gamma$, since they are qualitatively identical for the other two choices of $\omega\ts{r}$.
The central observation from Figs.~\ref{fig:lasing_scan}(a)-\ref{fig:lasing_scan}(d) is that the laser offset $\delta_0$ is much smaller than the corresponding detuning $\Delta$ for all parameters. Mathematically, this means that the slope of the offset's dependency on the detuning (cavity pulling coefficient) is smaller than one. For a conventional laser in the good cavity regime the cavity pulling coefficient is approximately one. In our case it is roughly between $0.1$ and $0.2$, depending on the pump rate $R$. In addition, the linewidth $\gamma$ does not vary much with the detuning. The significance of these two features is that the spectrum of the superradiant laser depends less on cavity fluctuations than the spectrum of a normal laser, which is the expected behaviour. Between $\Delta = 5\Gamma$ and $ \Delta = 15 \Gamma$ the frequency offset grows as expected, but for larger detunings it seems to reduce. The reason for this is that too large detunings lead to the formation of a distinct second peak approximately at the cavity resonance frequency, as shown in \fref{fig:spectrum_2peak}. Therefore, in these cases we determine the frequency close to the atomic resonance only, which almost vanishes.

The incoherent drive effectively broadens the atomic transition, resulting in an actual linewidth of $R+\Gamma$. We hence plot the FWHM in units of this effective atomic linewidth in \fref{fig:lasing_scan}(c) and \fref{fig:lasing_scan}(d). One can see that there are areas where the laser linewidth is even lower than the effective atomic linewidth. This is the case for high pump strengths and small detunings. The smallest laser linewidth, however, is achieved for small pump rates. We can also see that, for all stable parameters, the laser linewidth is well below the cavity linewidth, $\gamma < 2 \kappa$. The linewidth and the offset grow with increasing pump strength, which implies that the narrowest laser spectrum featuring a low frequency shift is achieved at small pump rates just above the lasing threshold. The atom-field coupling does not affect the offset, but the linewidth grows with it.

We note that the lasing properties shown in \fref{fig:lasing_scan} are almost identical for the three different $\omega\ts{r}$, which indicates that the laser properties do not change dramatically compared to a laser with fixed particle positions as described in \cite{maier2014superradiant}. To further support this statement we calculate the spectrum in the same manner as before, but for fixed atomic positions ($r_1 = 0$, $r_2 = \lambda_c/2$ and $r_3 = \lambda_c$). Comparing the resulting spectrum to the one with moving atoms, we find an almost perfect overlap. Therefore, the atomic motion appears to merely change the effective atom-field coupling, which does not significantly alter the spectrum.

\subsection{Photon number, second-order correlation and population inversion}
Besides the spectrum we also calculate other characteristic quantities of a laser. Specifically, we compute the average intra-cavity photon number,
\begin{equation}
n = \braket{a^{\dagger} a},
\label{eq:photon_number}
\end{equation}
and the second-order correlation function at zero time delay,
\begin{equation}
g^{(2)}(0) = \frac{\braket{a^{\dagger}a^{\dagger}aa}}{\braket{a^{\dagger}a}^2}.
\label{eq:2nd_order_cor}
\end{equation}
Finally, the population inversion of the atoms is relevant as well. The overall excitation is given by
\begin{equation}
p\ts{e} = \sum_{i=1}^N \braket{\sigma^+_i \sigma^-_i}
\label{eq:population}
\end{equation}
where inversion is achieved if $p\ts{e}>N/2$.

Figs.~\ref{fig:lasing_scan}(e)-\ref{fig:lasing_scan}(h) depict $n$ and $g^{(2)}(0)$ as functions of the scan parameters for $\omega\ts{r} = 1 \Gamma$. The most significant feature is that we always have less than half a photon on average inside the cavity. The figure also shows that the most photons are created and the field is most coherent ($g^{(2)}(0) = 1$) for small detunings, large pump strengths and large atom-field coupling. This behaviour coincides with that of a conventional (good-cavity) laser. The excited state population is always above $1.5$, and we note that the overall scan of the atomic excitation is similar to that obtained for a conventional laser.

Comparing the cooling and the lasing scan, we find that the optimal lasing point does not coincide with the best cooling, specifically for the pump strength dependency. We therefore conclude that there is a certain trade-off between the optimal cooling and lasing regimes.

\section{Conclusions}
We have seen that even less than one average intra-cavity photon can be sufficient in order to accumulate excited state atoms dynamically at positions of maximal light coupling, i.e. at field mode antinodes, in the blue-detuned regime. For a sufficient pumping one can thus achieve population inversion and gain which subsequently leads to superradiant lasing. This behaviour is stable with respect to forces and heating induced by dipole-dipole interaction. The output spectrum of such a laser exhibits a very low sensitivity to cavity length fluctuations with a linewidth determined by the atomic linewidth broadened by the pump rate. We have obtained these results by means of a semiclassical model, in which we have treated the atomic states as well as the cavity field mode quantum mechanically, whereas the atomic motion has been described classically.

Overall, for sufficiently slow atoms, the atomic motion only marginally affects the operating conditions and output characteristics of such a laser. In particular, its spectral and coherence properties remain almost unchanged as long as the photon numer is low. This is a promising result for the construction of a superradiant laser, where inverted atoms are moved through the cavity by an optical lattice conveyer belt. It seems that using as many atoms as possible with a weak pump and a large bandwidth cavity is the optimal way to operate such a device. Note, that we have used a rather generic rate based spatially uniform pumping scheme. This should be refined and modeled in more detail for future considerations.

\section*{Funding}
This project has received funding from the European Union's Horizon 2020 research and innovation programme under grant agreement No 820404 (iqClock) (C.~H. and H.~R.) and from the Austrian Science Fund (FWF) through projects DK-ALM W1259-N27 (D.~P.) and P29318-N27 (L.~O.).

\section*{Acknowledgments}
The numerical simulations were performed with the open source framework QuantumOptics.jl~\cite{kramer2018quantumoptics}.

\appendixtitleon
\appendixtitletocon
\begin{appendices}
\section{Semiclassical master equation for dipole-dipole interacting atoms} \label{app:semi_class_approx} 
In the following we develop the semiclassical description of our model. The internal atomic degrees of freedom as well as the cavity mode will be described in a quantum mechanical sense, whereas the atomic motion will be written in terms of classical variables only. We start from a full quantum model describing the coupling of moving two-level atoms to a cavity mode as well as to a continuum of free space vacuum modes. The Hamiltonian reads
\begin{equation} \label{eq:total_hamiltonian_appendix}
\begin{aligned}
H\ts{tot} &= H_0 + \sum_i\hbar g(\hat{r}_i^{\text{c}})\left(a^\dag \sigma_i^- + \sigma_i^+a\right) + \sum_{\textbf{k},\lambda}\hbar\omega_k b_{\textbf{k},\lambda}^\dag b_{\textbf{k},\lambda}
\\
&+ \sum_i \sum_{\textbf{k}, \lambda}\hbar g_{\textbf{k},\lambda}\left( b_{\textbf{k},\lambda}^\dag \sigma_i^- e^{-i\textbf{k}\cdot \hat{\textbf{r}}_i} + \text{H.c.}\right) + \sum_i \frac{\hat{\textbf{p}}_i^2}{2m},
\end{aligned}
\end{equation}
where the modes of the free space vacuum are described by the bosonic creation and annihilation operators, $b_{\textbf{k},\lambda}^\dag$ and $b_{\textbf{k},\lambda}$, respectively. Each wavevector $\textbf{k}$ features two polarizations $\lambda=1,2$. The (generally 3D) motion of the atoms is accounted for by the position and momentum operators $\hat{\textbf{r}}_i$ and $\hat{\textbf{p}}_i$. The coupling to the cavity mode is determined by the component of the position along the cavity axis $\hat{r}_i^{\text{c}}= \textbf{k}\ts{c}\cdot \hat{\textbf{r}}_i/k\ts{c}$. The free energy part of the cavity and the atoms is given by
\begin{align}
H_0 &:= \hbar\omega\ts{c}a^\dag a + \hbar \omega\ts{a}\sum_i \sigma_i^+\sigma_i^-
\end{align}
Note, that we perform the so-called independent bath assumption for the atoms and the cavity, i.e., the cavity decay does not affect the coupling of the atoms to the environment. Since the cavity damping does not affect the motion of the atoms directly either, we will neglect it for now.

The density operator describing the internal atomic dynamics as well as the motional degrees of freedom, the cavity mode and the 3D vacuum modes $\rho\ts{tot}$ is then governed by the von Neumann equation,
\begin{align}
\dot{\rho}\ts{tot} &= -\frac{i}{\hbar}[H\ts{tot},\rho\ts{tot}].
\end{align}
Essentially, the semiclassical approximation consists of two assumptions. First, we assume that there are no correlations (entanglement) between the motion and the remaining degrees of freedom. Secondly, we will assume that the motion is classical such that all expectation values factorize. The assumption that there are no correlations between the motion and the remaining degrees of freedom amounts to setting
\begin{align} \label{eq:rho_tot}
\rho\ts{tot}(t) &\approx \rho\ts{acf}(t)\otimes \rho\ts{m}(t).
\end{align}
On the one hand, the density operator $\rho\ts{acf}$ describes the state of the atomic excitation, the cavity, as well as the free-space vacuum modes. On the other hand, the motional degrees of freedom are given by $\rho\ts{m}(t)$. We now aim at finding an equation for the reduced system density operator $\rho\ts{acf}$. To this end, we take the partial trace,
\begin{align}
\dot{\rho}\ts{acf} &= -\frac{i}{\hbar}\tr\ts{m}\left([H\ts{tot},\rho\ts{tot}]\right) = -\frac{i}{\hbar}[H\ts{acf},\rho\ts{acf}].
\end{align}
Here, we have defined the reduced Hamiltonian
\begin{equation}
\begin{aligned}
H\ts{acf} &:= H_0 + \sum_i g(r_i^{c}(t))\left(a^\dag\sigma_i^- + \sigma_i^+ a\right) + \sum_{\textbf{k},\lambda}\hbar\omega_k b_{\textbf{k},\lambda}^\dag b_{\textbf{k},\lambda}
\\
&+ \sum_i \sum_{ \textbf{k},\lambda}\hbar g_{\textbf{k},\lambda}\left(b_{\textbf{k},\lambda}^\dag\sigma_i^- e^{-i\textbf{k}\cdot \textbf{r}_i(t)} + \text{H.c.}\right),
\end{aligned}
\end{equation}
where we wrote
\begin{align}
\textbf{r}_i(t) &= \braket{\hat{\textbf{r}}_i}(t) = \tr\left(\hat{\textbf{r}}_i\rho\ts{m}(t)\right).
\end{align}
Additionally, we made our second assumption of treating the motion clasically, such that $\braket{f(\hat{\textbf{r}})}\approx f(\textbf{r})$ for any function $f$.

The assumptions from  above constitute our semiclassical approximation. We can now proceed by eliminating the field modes. This leads to the dipole-dipole interactions among the atoms in the form of coherent energy exchange as well as collective decay. The only additional assumption one has to make to arrive at this is the Markov approximation for the atomic positions. The remaining procedure remains the same and yields~\cite{lehmberg1970radiation},
\begin{align}
\dot{\rho} &= \tr\ts{f}\left(\dot{\rho}\ts{acf}\right) = -\frac{i}{\hbar}[H,\rho] + \mathcal{L}\ts{cd}[\rho],
\end{align}
with $H$ the Hamiltonian from \eqref{Hamiltonian}.

The motion of the atoms, however, is still determined by $H\ts{tot}$ via 
\begin{equation}
\dot{\textbf{p}}_i = \tr(\hat{\textbf{p}}_i \dot{\rho}\ts{tot}) = - \frac{i}{\hbar} \tr(\hat{\textbf{p}}_i [H\ts{tot}, \rho\ts{tot}])
\end{equation}
and equivalently for $\dot{\textbf{r}}_i$. While the velocity is given in \eqref{ehrenfest_dr}, for the average force on the $i$th particle we have (in 1D)
\begin{align}
\dot{p}_i &= -\hbar \partial_{r_i} \sum_{j:j \neq i} \left(2\Omega_{ij}\Re{\braket{\sigma_i^+\sigma_j^-}} + \Gamma_{ij}\Im{\braket{\sigma_i^+\sigma_j^-}}\right).
\end{align}
Note that the term proportional to the collective decay does not significantly contribute since in our system $\Im{\braket{\sigma_i^+\sigma_j^-}}\approx 0~\forall~i,j$ due to almost perfect phase invariance~\cite{meiser2009prospects}.

\section{Calculation of the spectrum} \label{app:calc_spec}
The spectrum is given by the Fourier transform of the correlation function
\begin{align}
g(\tau) &= \braket{a^\dag(t+\tau) a(t)}.
\end{align}
A common method to calculate this is to define a new density operator at time $t$ which is then evolved up to a time $t+\tau$. This is also known as the optical regression theorem and the essential steps are as follows. Let $H\ts{tot}$ be a Hamiltonian which describes the entire system and bath dynamics. The evolution of the system is then reversible and given by the unitary operator $U(t)=\exp(-iH\ts{tot}t/\hbar)$. Thus, we can write the correlation function as
\begin{align}
g(\tau) = \tr\left(U^\dag(t)U^\dag(\tau)a^\dag U(\tau) a U(t)\rho\ts{tot}(0)\right) = \tr\left(a^\dag U(\tau) a \rho\ts{tot} U^\dag(\tau)\right),
\end{align}
where $\rho\ts{tot}$ is the total density operator. Upon defining a new density operator $\bar{\rho}\ts{tot}(0) := a \rho\ts{tot}(t)$, we may write
\begin{align}
g(\tau) = \tr\left(a^\dag\bar{\rho}\ts{tot}(\tau)\right).
\end{align}

The time evolution of $\bar{\rho}\ts{tot}$ is given by the same unitary operator as before. Therefore, eliminating the bath leads to the same master equation, but for a new operator $\bar{\rho}=a\rho$. In this way it is possible to compute $g(\tau)$ from the reduced system density operator via the master equation.

One has to be careful when deriving the semiclassical master equation for $\bar{\rho}$, though. In particular, the force on the atoms is proportional to average values of system variables such as $\braket{a^\dag\sigma_j}$. These have to be computed from the actual density operator $\rho$, rather than from $\bar{\rho}$. Assuming, as before, that there is no entanglement between the atomic motion and the remaining degrees of freedom [\eqref{eq:rho_tot}] is equivalent to writing $U(t)\approx U\ts{acf}(t)\otimes U\ts{m}(t)$. We can hence write
\begin{align} \label{eq:rho_bar_tau}
\bar{\rho}\ts{tot}(\tau) = U\ts{acf}(\tau)a\rho\ts{acf}(t) U\ts{acf}^\dag(\tau) \otimes U\ts{m}(\tau)\rho\ts{m}(t)U\ts{m}^\dag(\tau) = \bar{\rho}\ts{acf}(\tau) \otimes \rho\ts{m}(t+\tau),
\end{align}
where in the second step we have implicitly defined $\bar{\rho}\ts{acf}(0) := a \rho\ts{acf}(t)$ and have used the fact that $a$ does not act on the motional degrees of freedom. It is then possible to obtain a semiclassical master equation for $\bar{\rho}$ by tracing out the motion as well as the vacuum modes. However, as can be seen from \eqref{eq:rho_bar_tau}, the motional degrees of freedom are still determined by $\rho\ts{m}(t+\tau)$. Thus, we need to compute the motion up to a time $t+\tau$ in a time evolution with the density operator $\rho$. Only then can we calculate the proper time evolution of $\bar{\rho}$ by using the previously calculated particle positions and obtain the correlation function $g(\tau)$.

\begin{figure}[t]
\centering
\centering\includegraphics[width=0.8\columnwidth]{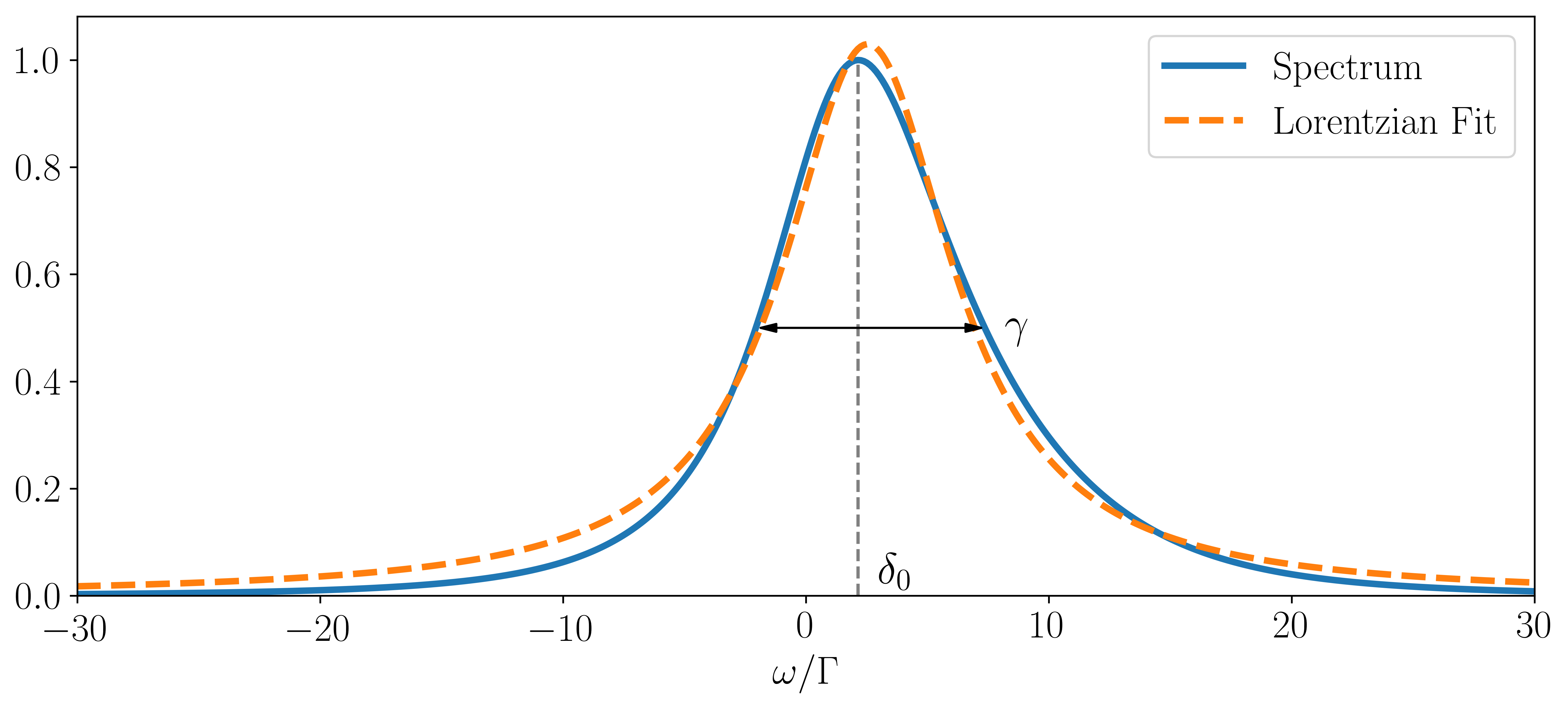}
\caption{\emph{Lorentzian fit of the normalized spectrum}. The calculated spectrum is fitted with a three-parameter Lorentzian function. We call the FWHM $\gamma$ and the offset to the atomic resonance frequency $\delta_0$. The parameters are the same as in \fref{fig:position_ab1_ab22_ekin_average}(b).}
\label{fig:spectrum_fit}
\end{figure}

If the detuning between the cavity and the atoms is not too large, the cavity output spectrum can be well described by a Lorentzian distribution, see \fref{fig:spectrum_fit}.

\section{Stability} \label{app:stability}
\begin{figure}[h]
\centering
\centering\includegraphics[width=0.8\columnwidth]{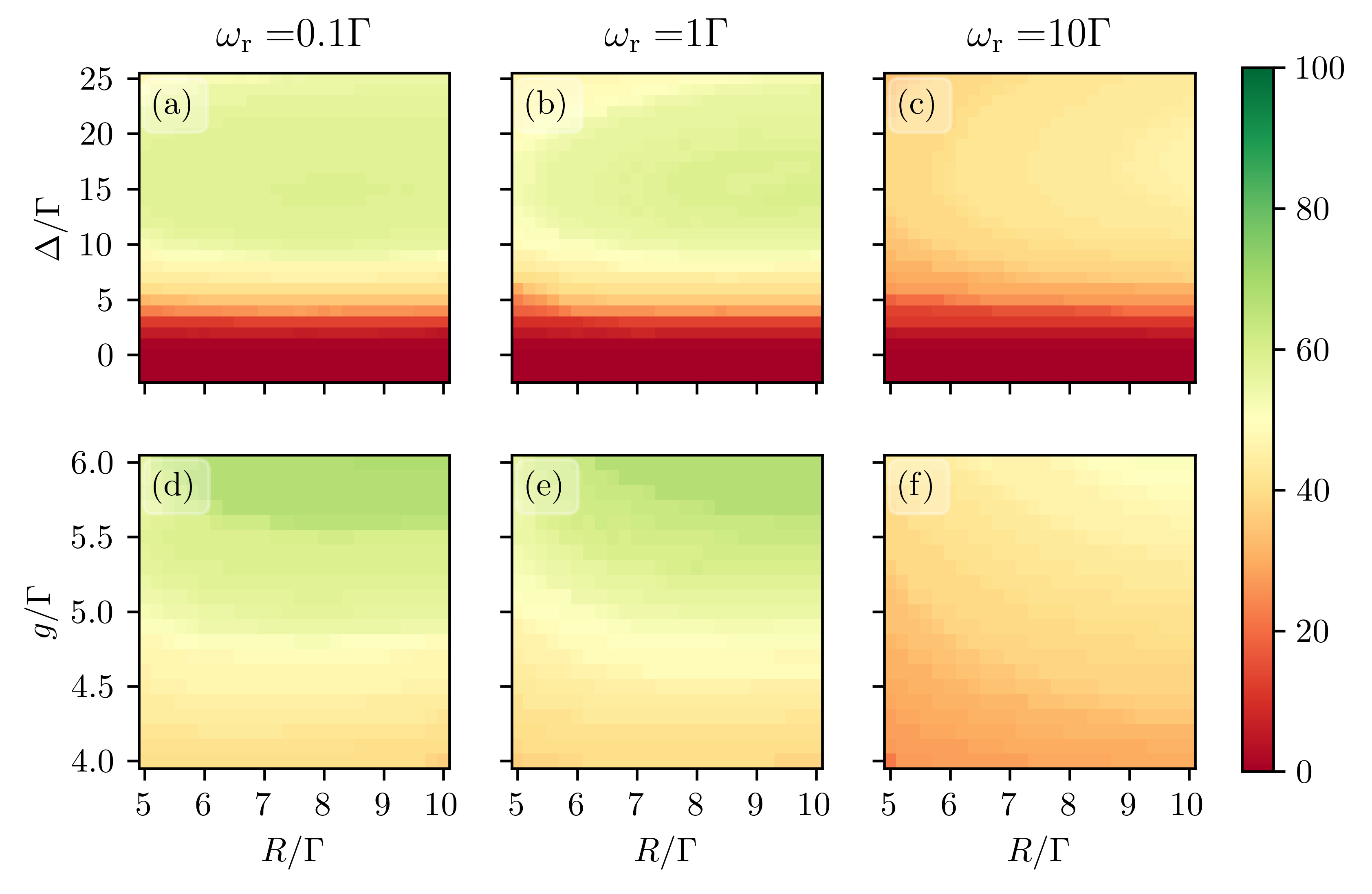}
\caption{\emph{Percentage of completely stable trajectories}. For $\Delta \leq 0$ there are no stable trajectories, because the atoms are heated and leave their initial trap. The fixed parameters are the same as in \fref{fig:e_kin_rel_scan}.}
\label{fig:stability}
\end{figure}

In \fref{fig:stability} we provide a scan of the number of completely stable trajectories, i.e.\ trajectories where the atoms stay in their initial trap for the whole time evolution. 
As mentioned in Sec. \ref{sec:cooling_and_trapping_properties} we see that lighter particles with the same kinetic energy are more difficult to trap. This is simply because they have a higher initial velocity and therefore it is harder for the cavity field to keep them trapped. Also, it takes a certain amount of time for the cavity to build up a sufficiently strong field which can confine the atoms. If the atomic velocity is too large, an atom may leave its initial trap during this build-up time.
\end{appendices}

\newpage
\bibliography{LaserRef}

\end{document}